\documentclass[draft]{agujournal}
\draftfalse

\usepackage[version=4]{mhchem}
\usepackage{amssymb,siunitx,amsmath,astjnlabbrev}
\journalname{Geophysical Research Letters}

\begin{document}

%% ------------------------------------------------------------------------ %%
%  Title
%
% (A title should be specific, informative, and brief. Use
% abbreviations only if they are defined in the abstract. Titles that
% start with general keywords then specific terms are optimized in
% searches)
%
%% ------------------------------------------------------------------------ %%

% Example: \title{This is a test title}

\title{Magnetospheric Multi-Scale  Observations of High-Energy Electrons and Protons in the Vicinity of Southern High-Altitude Cusp Boundaries}

%% ------------------------------------------------------------------------ %%
%
%  AUTHORS AND AFFILIATIONS
%
%% ------------------------------------------------------------------------ %%

% Authors are individuals who have significantly contributed to the
% research and preparation of the article. Group authors are allowed, if
% each author in the group is separately identified in an appendix.)

% List authors by first name or initial followed by last name and
% separated by commas. Use \affil{} to number affiliations, and
% \thanks{} for author notes.
% Additional author notes should be indicated with \thanks{} (for
% example, for current addresses).

% Example: \authors{A. B. Author\affil{1}\thanks{Current address, Antartica}, B. C. Author\affil{2,3}, and D. E.
% Author\affil{3,4}\thanks{Also funded by Monsanto.}}

\authors{K. Nykyri \affil{1}, C. Chu \affil{1}, X. Ma \affil{1}, R. Rice \affil{1}}

% \affiliation{1}{First Affiliation}
% \affiliation{2}{Second Affiliation}
% \affiliation{3}{Third Affiliation}
% \affiliation{4}{Fourth Affiliation}

\affiliation{1}{Physical Sciences Department  and Centre for Space and Atmospheric Research, Embry-Riddle Aeronautical University, 600. S. Clyde Morris Blvd., Daytona Beach, FL, USA}
%(repeat as many times as is necessary)

%% Corresponding Author:
% Corresponding author mailing address and e-mail address:

% (include name and email addresses of the corresponding author.  More
% than one corresponding author is allowed in this LaTeX file and for
% publication; but only one corresponding author is allowed in our
% editorial system.)

% Example: \correspondingauthor{First and Last Name}{email@address.edu}

\correspondingauthor{Katariina Nykyri}{nykyrik@erau.edu}

%% Keypoints, final entry on title page.

% Example:
% \begin{keypoints}
% \item	List up to three key points (at least one is required)
% \item	Key Points summarize the main points and conclusions of the article
% \item	Each must be 100 characters or less with no special characters or punctuation
% \end{keypoints}

%  List up to three key points (at least one is required)
%  Key Points summarize the main points and conclusions of the article
%  Each must be 100 characters or less with no special characters or punctuation

\begin{keypoints}
\item MMS orbit can reach southern high-altitude cusp and associated boundaries.
 \item MMS observed flows and magnetic field rotation consistent with magnetic reconnection occurring dawn-ward and above of the MMS spacecraft.  
 \item MMS observed closely 90 degree pitch angle high-energy electrons and protons in the depressed magnetic field regions with nearly stagnated flow.
%\item Diamagnetic cavities may be the source for some of the high-energy particles observed in the magnetosheath

\end{keypoints}

%% ------------------------------------------------------------------------ %%
%
%  ABSTRACT
%
% A good abstract will begin with a short description of the problem
% being addressed, briefly describe the new data or analyses, then
% briefly states the main conclusion(s) and how they are supported and
% uncertainties.
%% ------------------------------------------------------------------------ %%

%% \begin{abstract} starts the second page

\begin{abstract}
Recent Magnetosphere Multiscale (MMS) observations found 238 high energy
($>$ 40 keV) electron ``leaking'' events in the the magnetosheath {\it\citep{cohen17}}. While several sources have been proposed, the dominant mechanism or origin of these particles is not well understood. We have analyzed MMS locations during these these events and found that most of these electron leaking events were observed close to the southern high-altitude cusp and associated boundaries, so these events may have a high-latitude source.  Here we present a new case study of observations of a MMS encounter of southern magnetospheric cusp boundaries. In the magnetosheath side of the southern cusp,  MMS observes both parallel streaming energetic ions, as well as a population with 90 degree pitch angles. Subsequently, MMS detects plasma flows and magnetic field rotation consistent with magnetic reconnection (operating above and dawn-ward of the MMS spacecraft), finally entering a depressed magnetic field region with nearly stagnant flow.  The depressed field region is occupied by high-fluxes of trapped high-energy electrons and protons, which resemble the Cluster observations of cusp diamagnetic cavities at the northern hemisphere \citep{nykyo11a}. However, the local magnetic field geometry during the prevailing IMF orientation is not favorable for magnetic reconnection to directly create a cavity in this location. We show that the plasma flows that were observed prior to cavity observation satisfy the onset condition for the Kelvin-Helmholtz instability for 50 percent of all available $k$-vector directions.
\end{abstract}

%% ------------------------------------------------------------------------ %%
%
%  TEXT
%
%% ------------------------------------------------------------------------ %%

%%% Suggested section heads:
% \section{Introduction}
%
% The main text should start with an introduction. Except for short
% manuscripts (such as comments and replies), the text should be divided
% into sections, each with its own heading.

% Headings should be sentence fragments and do not begin with a
% lowercase letter or number. Examples of good headings are:

% \section{Materials and Methods}
% Here is text on Materials and Methods.
%
% \subsection{A descriptive heading about methods}
% More about Methods.
%
% \section{Data} (Or section title might be a descriptive heading about data)
%
% \section{Results} (Or section title might be a descriptive heading about the
% results)
%
% \section{Conclusions}

\section{Introduction}

%xuan: I think you can put this sentence to the end of this paragraph.
 Recent Magnetosphere Multi-Scale (MMS) observations have revealed high energy ($\ge$ 40 keV) electrons leaking into the magnetosheath {\it\citep{cohen17}}. While the Global Lyon-Fedder-Mobarry (LFM) MHD model {\it\citep{Sorathia2017}} with test particles suggest that magnetic reconnection and the nonlinear Kelvin-Helmholtz instability (KHI) could cause the leakage of high energy electrons into the magnetosheath, the origin and the dominant leaking mechanism is not well understood.  It is important to note that many of the electron leaking events were observed close to Equinoxes when the MMS orbit has a significant $y$-component and the $z_{GSM}$ coordinate can be substantial (up to $\approx $ 5-7 R$_{E}$).  As an example, Figure 1 shows the MMS location in GSM coordinates for six of the observed electron leakage events (one from each month of the list provided by {\it\citet{cohen17}}). It is evident that closer to the fall (23rd of September, 2015) and spring (March 20th) equinoxes, the MMS orbit has large $z_{GSM}$-coordinates and is close to the southern cusp or associated boundaries. The November and December events, while having a small $z_{GSM}$-coordinate, are close to the southern dayside magnetopause and possibly the cusp boundary when dayside magnetosphere is tilted toward positive $z_{GSM}$. It therefore may be possible that some of the MMS high-energy electron events observed in the magnetosheath may indeed have a high-latitude source. 
   
 It has been well demonstrated that magnetic reconnection between the Interplanetary Magnetic Field (IMF) and Earth's magnetic field surrounding the magnetospheric cusps can lead to the formation of cusp DiaMagnetic Cavities (DMCs) {\it\citep{nykyo11a,nykyo11b,adamso11,adamso12}}, extended regions of decreased magnetic field surrounding the high-altitude cusp funnel, which can be filled with higher energy ($>$ 30 keV) electrons, protons and \ce{O+} ions \citep{chenf98,fritz99,chenf01,zhanf05,whitc06,whitf07,walshf07,niehf08,walshf10,nykyo12}.  While the origin of these high-energy particles in the cavity has been under some debate (see e.g \citet{nykyo11a} and references therein), the test particle simulations \citep{nykyo12},  and the presence of the high fluxes of energetic 90 degree pitch-angle electrons in strongly depressed magnetic field regions  \citep{walshf10,nykyo11a,nykyo12} suggest that the bow shock or magnetosphere can not be the direct and the only source.  

Recently, \citet{luo17} performed a statistical study using 11 years of high energy ($>$ 274 keV) proton and oxygen data. Their results indicate that the energetic ion distributions are influenced by the dawn-dusk interplanetary magnetic field (IMF) direction.  Under northward IMF their statistics for high latitudes at 4 $R_E$ $< |Z | < 8$ $R_E$ showed a higher asymmetry for quadrants where the location of a diamagnetic cavity is predicted. During southward IMF with positive $B_y$ it was found that the intensity of H+ is much higher at the dusk-side than that at the dawn-side for both the dayside magnetosphere and nightside plasma sheet in the northern hemisphere.  However, this dusk-side favored asymmetry was absent in the Southern Hemisphere at the dayside.  Their study concluded that the location of the cusp diamagnetic cavities \citep{nykyo11a}, which depend on the IMF orientation,  can contribute to the observed asymmetry in the energetic ion population in the dayside magnetosphere and plasma sheet.

%Xuan: This paragraph sounds too long, and I can't catch the main point of this paragraph.

Polar and Cluster spacecraft observations have revealed that depending on the orbit altitude and prevailing solar wind conditions cusp crossings can look quite different when comparing magnetic field and plasma signatures. The outbound (northern hemisphere) cusp crossings from Cluster have revealed that reconnection tailward of the cusp during northward IMF leads into strong field aligned flows which are observed when spacecraft enter the reconnected cusp fields lines from tail lobe \citep{vontb03}. The magnetic field strength during these type of cusp crossings is still large, $\approx $ 100- 60 nT,  and gradually decreases to $\approx$ 40 nT. However, the de-trended magnetic field data has revealed strong fluctuations of the magnetic field and the existence of discrete wave modes and turbulence \citep{nykyc04,nykyg06}.  When moving further away from field lines that have recently reconnected into the region of accumulated by old reconnected flux, the spacecraft observe stagnant plasma. \citet{lavrd02} coined the term ``Stagnant Exterior Cusp (SEC)''  characterized by stagnant plasma and more isotropic ion velocity distributions.  
A statistical study of the properties of the 40 SEC  crossings has shown that the large amplitude magnetic field fluctuations are closely associated with the larger magnetic shear angle, which is consistent with high-latitude reconnection process \citep{zhanf05}. This study also found energetic  ($>$ 28 keV) protons during 80 percent, and energetic electrons during 23 percent of the SEC crossings.  

The magnetic field during SEC observed by  \citet{lavrd02} has shown a gradual decrease from $\approx$ 40 nT to $\approx$ 10 nT. The encounters of the DMCs, however, have revealed very abruptly and strongly depressed magnetic field with respect to surrounding boundaries. For example, four Cluster spacecraft had encounters with two DMCs during February 14th 2003, first during northward and subsequently during southward IMF (see Figure 1a).  The 5000 km spacecraft separation allowed for the first time a detailed determination of DMC structure and dynamics \citep{nykyo11a}. During the 1st cavity encounter the magnetic field rapidly dropped from 80 nT (in lobe magnetosphere)  to 4 nT (in cavity) at Cluster 1 which had the highest z-coordinate.  When IMF turned southward,  a new cavity formed sunward of the old cavity.  The 2nd cavity observations were bounded by observations of magnetosheath plasma and characterized  by magnetic field decrease from 60 nT (in the magnetosheath)  to $\approx$ 5 nT in the cavity. Throughout both cavity encounters strong magnetic field fluctuations were present. \citet{nykyo11b} showed that most of these fluctuations in the magnetic field strength were due to motion of the cusp boundaries or transient reconnection signatures. \citet{nykyo11a} reported that according to their survey Cluster encountered clear DMCs only when the dynamic pressure of the solar wind was high enough (typically above $\approx$ 2 nPa). During many (about one third) of the high-altitude cusp crossings Cluster observes a magnetic field that is not depressed like during DMCs but gradually decreases from $\approx$100 nT to $\approx$20 nT.  The Polar spacecraft observed DMCs during high-altitude cusp crossing for extended time periods, because the DMCs are in the apogee of the orbit and Polar moves very slowly through this region. Cluster moves faster through this region and at lower altitude, so it only encounters clear DMCs during intervals of enhanced dynamic pressure. For example, the inbound southern cusp crossing under southward IMF by Cluster (see Figure 1d) reported by \citet{cargd04} occurred during high SW dynamic pressure of $\approx$ 2-3 nPa, and the crossing shows a rapid depression of about 60 nT in the magnetic field strength lasting only about 5 minutes and coinciding with higher ion temperatures and reduced densities with respect to surrounding regions. 

In this paper we present a case study of Magnetosphere Multi-Scale (MMS) observations during October 2nd 2015, when MMS traversed dusk-ward from the dayside magnetosphere through the high-latitude dayside boundary layer, and had multiple encounters with trapped high-energy particle populations in depressed magnetic field regions. The duration of quasi- periodic encounters with the high energy particles lasted for several hours, but in this paper we focus on a sub-interval from 9:18-9:30 UT. The MMS trajectory between 8:00-11 UT is marked with red trace in Figures 1j and k. The MMS separation is only about 20-30 km, so all spacecraft observe essentially the same large scale plasma and magnetic field features. 

\section{Methods}
\subsection{Instrumentation and data used}
All magnetospheric data shown in Figure 2 is taken from NASA's four MMS satellites {\it\citep{burch2016}}. We use the level 2 data from the Fast Plasma Investigation (FPI) {\it\citep{pollock2016}} for the ion energy spectra and moments; Flux Gate Magnetometers (FGM) {\it\citep{russell16,torbert2014}} for the DC magnetic field, energetic electron distribution and pitch angle data will come from the Fly's Eye Energetic Particle Spectrometer (FEEPS) {\it\citep{Blake2016}} instrument. Proton and oxygen distribution and pitch angle data is available from the Energetic Ion Spectrometer (EIS) {\it\citep{Mauk2016}}. The versions of the data files used are v4.18.0.cdf,  v3.1.0.cdf, v6.0.1.cdf, and  v3.0.0.cdf for FGM, FPI, FEEPS and EIS, respectively. Solar wind conditions are taken from the OMNI (http://omniweb.gsfc.nasa.gov/) database \citep{king2005}.

\subsection{Global MHD modeling using BATSRUS}
In order to put the MMS observations in the context of the magnetospheric boundaries, we have simulated the event from 08:00 to 11:00 UT using Solar Wind Modeling Framework (SWMF/BATSRUS \citep{toths05}) using 34.7M cells and 1/16 $R_E$ numerical resolution at the inner boundary. The run results and model settings can be found at NASA community coordinated modeling center (CCMC) (https://ccmc.gsfc.nasa.gov/results) with the following run ID: $Katariina\_Nykyri\_020918\_2$. 

\subsection{Loss cone pitch angle calculation}
The loss cone pitch angle, 
\begin{eqnarray}
\alpha= \arctan(\frac{1}{\sqrt{B_{msp}/B_{cavity} -1}}),
\end{eqnarray}
is calculated assuming the conservation of the particle energy and first adiabatic invariant and assuming that particles in the strong field region only have perpendicular kinetic energy.
The loss cone pitch angle calculation uses a constant magnetospheric field value of 45 nT.

\section{MMS observations of energetic particles at southern cusp diamagnetic cavity and MHD simulations}

Figure 2 presents MMS1 observations of plasma and magnetic field properties during October 2nd 2015 between 09:18-09:30 UT (see caption for details on the panels). MMS is located in the vicinity of he dayside dusk sector of the southern cusp and dayside magnetosphere ($R$ $\approx$[7.9, 6.4, -4.3])). The IMF (panel h) is steady southward ($B_z$ $\approx$ -6--7 nT) with a strong dusk-ward component ($B_y$ $\approx$ +6-7 nT ). The $B_x$ varies between -1 to +0.5 nT.  Solar wind velocity varies between 360-400 km/s, and density varies between 3.9-5.5/cc inducing dynamic pressure of the order between 1.1-1.5 nPa during the interval. 
Between 9:18-9:18:40 UT MMS is in the magnetosheath (msh, yellow highlighted column), characterized by low energy plasma (panel c) with lower temperatures and high ion densities (panel d). Between $\approx$ 9:19-9:21:15 it encounters gradually increasing strong tail-ward plasma flows (panel e) and magnetic field rotation (panel g). The plasma density reduces from magnetosheath values to about 6-11/cc and temperature slightly increases. Magnetic field strength shows about 30 s oscillations with about 10 nT amplitude, creating a wavy signature in ion-beta (panel f). Comparison of the observed magnetic field and plasma flow with the magnetic field topology in Figure 1j and k is consistent with the MMS trajectory from the magnetosheath through the rotational discontinuity where $B_x$ and $B_z$ first become more negative when MMS enters the magnetosheath side of the reconnected field line, and then gradually turn positive when MMS moves to the magnetospheric side of the reconnected field line. The $B_y$ is positive (panel g) both on draped IMF field lines and on Earth's magnetic field lines in this location as can be expected based on Tsyganenko 96 \citep{tsyga96} model (see Figure 1j) and  from global MHD model  (see Figure 3a and d). Figure 4 shows that during between 9:18:30-9:21:15 UT there exists an excellent deHoffmanTeller frame (slope $=$ 1 and correlation coefficient $=$ 0.94), and a good Wal{\'e}n relation  (slope $=$ -0.822 and cc. $=$ -0.94). The deHoffman Teller frame (HT) velocity is tailward, poleward and dusk-ward, roughly consistent with the direction of purple arrow in Figure 1k. 

The magnetic field strength (panel k) shows field depression of about 22 nT. The depressed field regions correlates with enhanced fluxes of high energy (70-1000 keV) electrons (panels  a and j) and protons (48-209 keV) (panels b and i).  In pitch angle plots (panels i and j), the black lines represent the boundary of the loss cone for the particles inside the cavity: assuming adiabatic particle motion the particles that have pitch angles between the black lines are trapped and cannot originate from the higher magnetic field region directly without some reprocessing. In particular,  the 70-1000 keV electrons seem to be well trapped in the depressed field regions. In the magnetosheath there exists parallel high energy proton fluxes. These protons close to magnetopause boundary could originate from quasi-parallel bow shock \citep{trattner11} from the northern hemisphere. At the end of the main cavity observation between 9:23-9:25 UT, in addition to trapped proton population, MMS observes also anti-parallel high-energy protons in the loss cone. Because main magnetic field is in the positive $y$-direction during this interval, it appears these particles are moving from dusk to dawn potentially leaking out from the cavity.

 It is interesting to consider whether these reconnection flows prior to cavity observation could originate from component reconnection \citep{fuselier11b} dawn-ward and northward of the MMS. Figure 3  shows high-resolution MHD global simulation results,  together with MMS1 location projected in each plane,  of the plasma and field properties of the dayside magnetosphere at 09:24 UT. The diamagnetic cavities, directly generated by magnetic reconnection in maximum magnetic shear regions in similar manner as described by \citet{nykyo11a,adamso11, adamso12},  are indicated by a strongly enhanced plasma beta tailward of the MMS at $x=$ 4 $R_E$ (a). The  Alfv{\'e}n Mach number at the MMS location ($x=$ $R_E$) is nearly one, suggesting that this region could be KHI unstable for certain $k$-vector orientations, but also that reconnection may be suppressed in this region by strong flow \citep{chen97} (b). The $y$-component of the current density ($J_y$) in $x,z$-plane with a cut at the $y=$0 shows that $J_y$ is enhanced at the extended region around dayside magnetopause potentially favoring component reconnection (c). The  magnetic field strength in $z,y$-plane with a cut at MMS location ($x=$ 7.9 $R_E$) shows that magnetic field, while lower than in the magnetosphere,  does not quite reach such as low values as 
observed by the MMS. We also produced cuts along the simulated MMS orbit of the magnetic field, plasma flow velocity, density and temperature (not shown). The range of density ($n$), temperature ($T$), velocity ($v_x$, $v_y$, $v_z$), magnetic field ( $b_x$, $b_y$, $b_z$ and $b_t$) variation between 9:18-9:30 UT are as follows:  $n=$ [4.3,9.4]/cc, $T=$[3.5,6.5]eV, $v_x=$[-130,-100] km/s, $v_y=$[80, 105] km/s, $v_z=$[-150, -110] km/s, $b_x=$[-9,-4] nT, $b_y=$[27, 38] nT, $b_z=$[4, 14] nT, and $b_T=$[29, 36] nT. This indicates that the virtual MMS does not observe the fast flows or magnetic field rotations,  similar to MMS.

\section{Conclusions and Discussion}

The main conclusions of this paper can be summarized as follows: We have shown using MMS data and Tsyganenko 96 and Global MHD modeling that 

1. MMS orbit can reach the southern high-altitude cusp and associated boundaries.

2. MMS observed flows and magnetic field rotation consistent with magnetic reconnection occurring dawn-ward and above of the MMS spacecraft.  

3. MMS observed closely 90 degree pitch angle high-energy electrons and protons in the depressed magnetic field regions with nearly stagnated flow.

A puzzle remains is that why the diamagnetic cavities (characterized by high plasma beta, depressed magnetic field, trapped high-energy particles) formed in the ``wrong'' location. It is not clear how the ring current or radiation belt particles could manifest themselves in these high-latitude pockets of depressed magnetic field. Previous Cluster observations and the IMF orientation for the present event are consistent with the global MHD simulations of the expected cavity locations (see Figure 3a, e and f) at the northern dusk-side  and and at the southern dawn-side of the major cusp funnel. The component reconnection at the dayside, while qualitatively consistent with the HT frame and Wal{\'e}n relation observations,  is not supported by the high-resolution MHD simulations which should well resolve the magnetic reconnection at the dayside magnetopause. 

It is interesting to note that, while the velocity in the cavity is almost zero, the strong flows (see pink shaded region in Figure 2)  %($\sim \SI{300}{\kilo\meter\per\second}$)$ 
($\sim $ 400 km/s) before cavity observation (hereafter labeled as Boundary Layer (BL)) are orientated along the (southern) poleward and tailward directions. A region with such a large shear flow, that is mostly perpendicular to the magnetic field, can be Kelvin-Helmholtz (KH) unstable.
The KH instability onset condition requires 
\begin{eqnarray}
Q=\alpha_{\mbox{\tiny{BL}}}\alpha_{\mbox{\tiny{cavity}}}[(\mathbf{V}_{\mbox{\tiny{BL}}}-\mathbf{V}_{\mbox{\tiny{cavity}}})\cdot\mathbf{k}]^2-\alpha_{\mbox{\tiny{BL}}}(\mathbf{V}_{A_{\mbox{\tiny{BL}}}}\cdot\mathbf{k})^2-\alpha_{\mbox{\tiny{cavity}}}(\mathbf{V}_{A_{\mbox{\tiny{cavity}}}}\cdot\mathbf{k})^2>0,
\end{eqnarray} where  $\alpha_{\mbox{\tiny{BL}}}=n_{\mbox{\tiny{BL}}}/(n_{\mbox{\tiny{BL}}}+n_{\mbox{\tiny{cavity}}})$, $\alpha_{\mbox{\tiny{cavity}}}=n_{\mbox{\tiny{BL}}}/(n_{\mbox{\tiny{BL}}}+n_{\mbox{\tiny{cavity}}})$, $n$ is the number density, $V_A=B\sqrt{\mu_0m_in}$ is the Alfv{\'e}n velocity, and the $\mathbf{k}$ is the KH wave-vector.
To examine whether the configuration between 9:21-9:24 UT is KH stable or not, we assume that the first half minute as the boundary layer side, the last half minute as the cavity side, which gives $n_{\mbox{\tiny{BL}}}=\SI{4.89}{cm^{-3}}$, $n_{\mbox{\tiny{cavity}}}=\SI{1.99}{cm^{-3}}$, $\mathbf{B}_{\mbox{\tiny{BL}}}= [11.18,31.33,18.12]\si{nT}$,
$\mathbf{B}_{\mbox{\tiny{cavity}}}= [ 6.53,   19.79,   15.44]\si{nT}$, $\mathbf{V}_{\mbox{\tiny{BL}}}= [-304.07, 59.57, -184.25]\si{km/s}$, and
$\mathbf{V}_{\mbox{\tiny{cavity}}}= [ -41.75,  -91.24,  -44.03]\si{km/s}$.
Figure 5 shows the value $Q$ as the function of the unit vector of the KH $k$-vector in  the spherical coordinates. The color code values above zero mark the $k$-vector orientations with respect to the velocity shear that satisfy the onset condition between 9:21-9:24 UT. The computation of the solid angle of the unstable $k$-vector orientations  indicates that about 50 percent of all the possible $k$-vector directions are KH unstable. It may therefore be possible that the   KH instability operating at the high-altitude southern cusp may contribute to the twisting of the local magnetic field close to the MMS spacecraft generating magnetic reconnection and creating a magnetic bottle configuration where local energization could be possible via particle drift in reconnection quasi-potential \citep{nykyo12}. Indeed, KH instability has been observed before at the high-altitudes \citep{hwang12,mxaopdzh15}. 

In conclusion,  the role of the magnetic reconnection and KH instability on the generation of the diamagnetic cusp cavities,  as well as the possible local acceleration in the vicinity of the high-latitude and altitude magnetospheric cusps needs to be further studied with help of 3-D global MHD simulations with test particles and with more MMS events. It is likely,  based on this case study and analysis of the MMS orbits during ``leakage events'' that some of the high energy electrons originate from these depressed magnetic field regions at high-altitude cusps with high fluxes of high-energy particles.

%Text here ===>>>

%%

%  Numbered lines in equations:
%  To add line numbers to lines in equations,
%  \begin{linenomath*}
%  \begin{equation}
%  \end{equation}
%  \end{linenomath*}

%% Enter Figures and Tables near as possible to where they are first mentioned:
%
% DO NOT USE \psfrag or \subfigure commands.
%
% Figure captions go below the figure.
% Table titles go above tables;  other caption information
%  should be placed in last line of the table, using
% \multicolumn2l{$^a$ This is a table note.}
%
%----------------
% EXAMPLE FIGURE
%

\begin{figure}[h]
\centering
% when using pdflatex, use pdf file:
%\includegraphics[width=40pc]{Final_figures/Figure_1new.pdf}
\includegraphics[width=40pc]{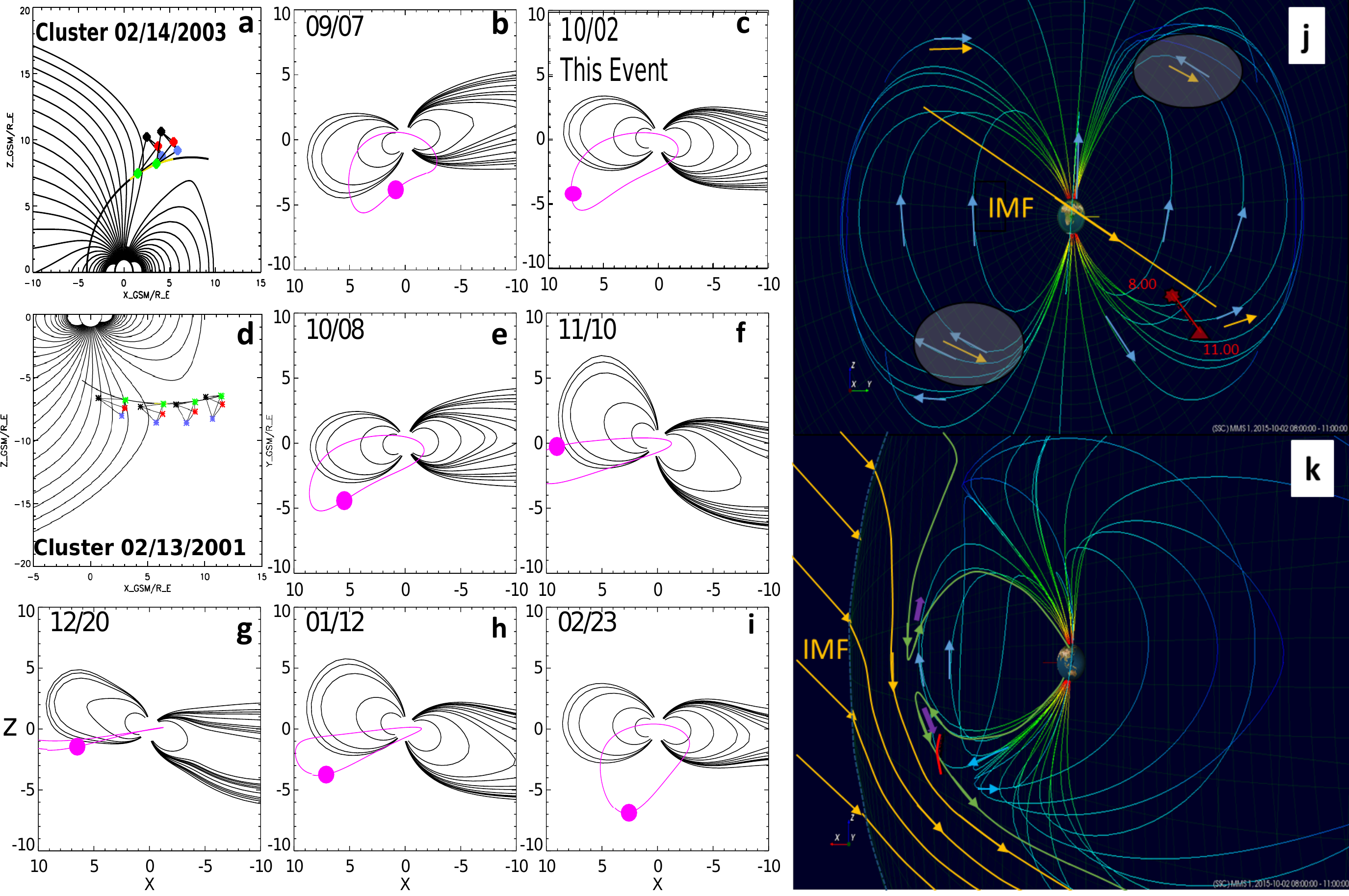}

% when using dvips, use .eps file:
% \includegraphics[width=20pc]{figsamp.eps}
%
 \caption{Examples of MMS locations close to southern cusp and associated dayside magnetospheric boundaries calculated from T96 \citep{tsyga96} model during  electron leakage events observed by \citet{cohen17} (b,e,f,g,h,i) and for the present study (c) on 10/02/2015 in GSM coordinates. Cluster trajectory during northern (a) and southern (d) cusp crossings from \citet{nykyo11a,cargd04}, respectively.  3-D visualization of MMS orbit using T96 model at 9:24 UT in $z-y$-plane (j) and in $x-z$-plane with view from positive $y$-axis (k). The MMS orbit between 8-11:00 UT is shown in red in panels j and k, and the orbit for the entire day is shown in magenta color in other plots. The IMF field lines are visualized in orange. 
 %The draping about IMF with strong negative $B_z$ ($\approx $ -6 - -7  nT) and $B_y$ ($\approx$ +6 - +7nT ) leads to anti-parallel magnetic fields at the post-noon sector of the northern cusp (see cavity iv) in Figure 3 e) and at pre-noon sector of the souther cusp (see cavity iv) in Figure 3f). (j) 
 The expected locations of anti-parallel reconnection and cavity formation are marked with shaded ovals. MMS observes flows (purple arrows) and magnetic field rotation (green vectors) that are consistent with the component reconnection originating possibly from more equatorial reconnection site above MMS.}
 \label{T96}
  \end{figure}

\begin{figure}[h]
\centering
% when using pdflatex, use pdf file:
\includegraphics[width=25pc]{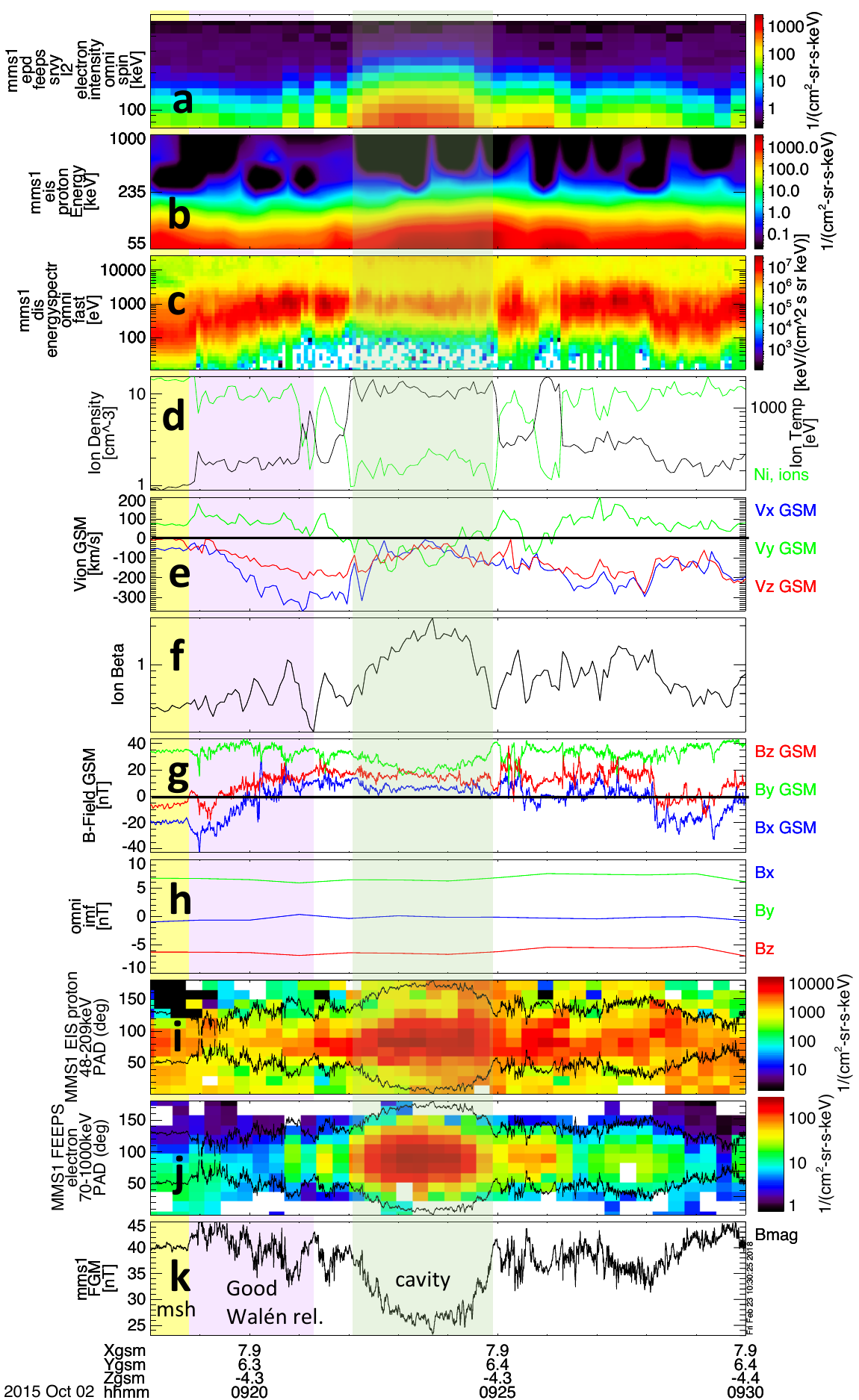}

% when using dvips, use .eps file:
% \includegraphics[width=20pc]{figsamp.eps}
%
 \caption{From top to bottom the panels show MMS1 observations of (a) energetic electrons (from FEEPS), (b) protons (from EIS), (c) lower energy protons (FPI), (d) ion density and temperature (FPI), (e) ion velocity (FPI), (f) ion plasma beta, (g) magnetic field, (h) OMNI magnetic field, (i) energetic ion and electron (j) pitch angle distributions and (k) total magnetic field strength. The areas outside black envelope in pitch angle plots shows the loss cone. The depressed magnetic field regions show trapped high energy electron and proton population. Magnetosheath (msh) region is marked with yellow, interval satisfying the Wal{\'e}n relation with pink, and cavity with green box. }
 \label{figone}
  \end{figure}

\begin{figure}[h]
\centering
% when using pdflatex, use pdf file:
\includegraphics[width=33pc]{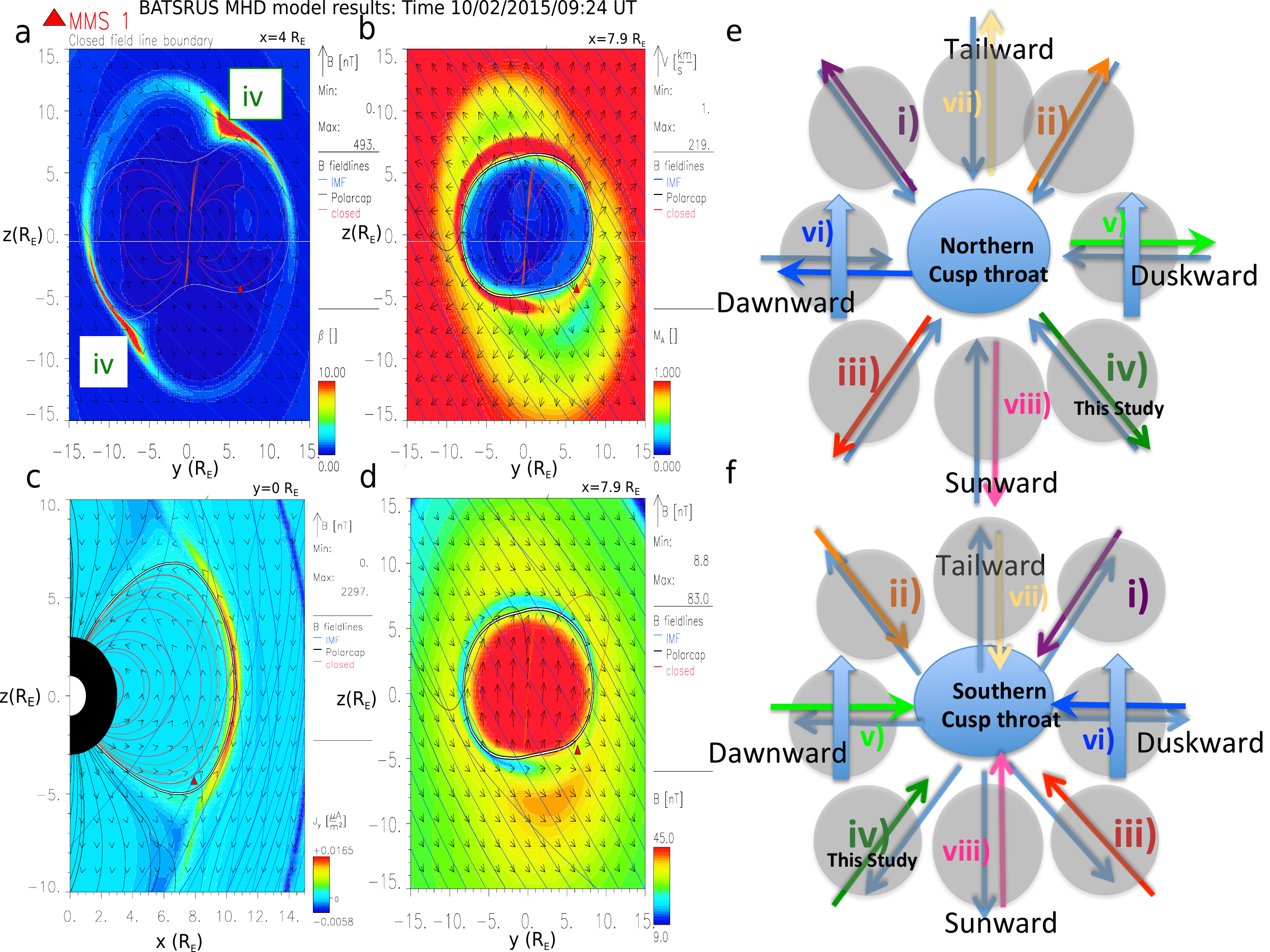}

% when using dvips, use .eps file:
% \includegraphics[width=20pc]{figsamp.eps}
%
 \caption{ BATSRUS MHD model results at 9:24 UT of plasma beta in $y-z$-plane at $x=$4 $R_E$ (a), Alfv{\'e}n Mach number at $x=$7.9 $R_E$ (b), magnetic field strength at $x=$7.9 $R_E$ (d), and current density in $x-z$-plane at $y=$0 (c). The MHD model resolves the diamagnetic cavities well, as indicated by enhanced plasma beta,  in post-noon sector of northern cusp (iv) and at pre-soon sector of the southern cusp (iv). Diagram of the expected locations (grey ovals) of the Cusp Diamagnetic Cavities for different IMF orientations at the northern (e) and southern (f) cusp. Grey arrows presented the direction of Earth's magnetic field and colored arrows present the direction of the draped IMF for the following main IMF $B_{y}$ and $B_{z}$ conditions: i) $B_y <$ 0, $B_z > $0, ii) $B_y >$ 0, $B_z >$0, iii) $B_y <$0, $B_z <$0, iv) $B_y >$ 0, $B_z <$0, v) $B_y \gg |B_z|$, vi) $B_y \ll-|B_z|$, vii) $B_z \gg |B_y|$, viii) $B_z \ll -|B_y|$. The blue thick arrows present regions where B-field is perpendicular the magnetosheath flow which can be high-latitude KH unstable.}
 \label{mhd_cartoon}
  \end{figure}

\begin{figure}[h]
\centering
% when using pdflatex, use pdf file:
\includegraphics[width=20pc]{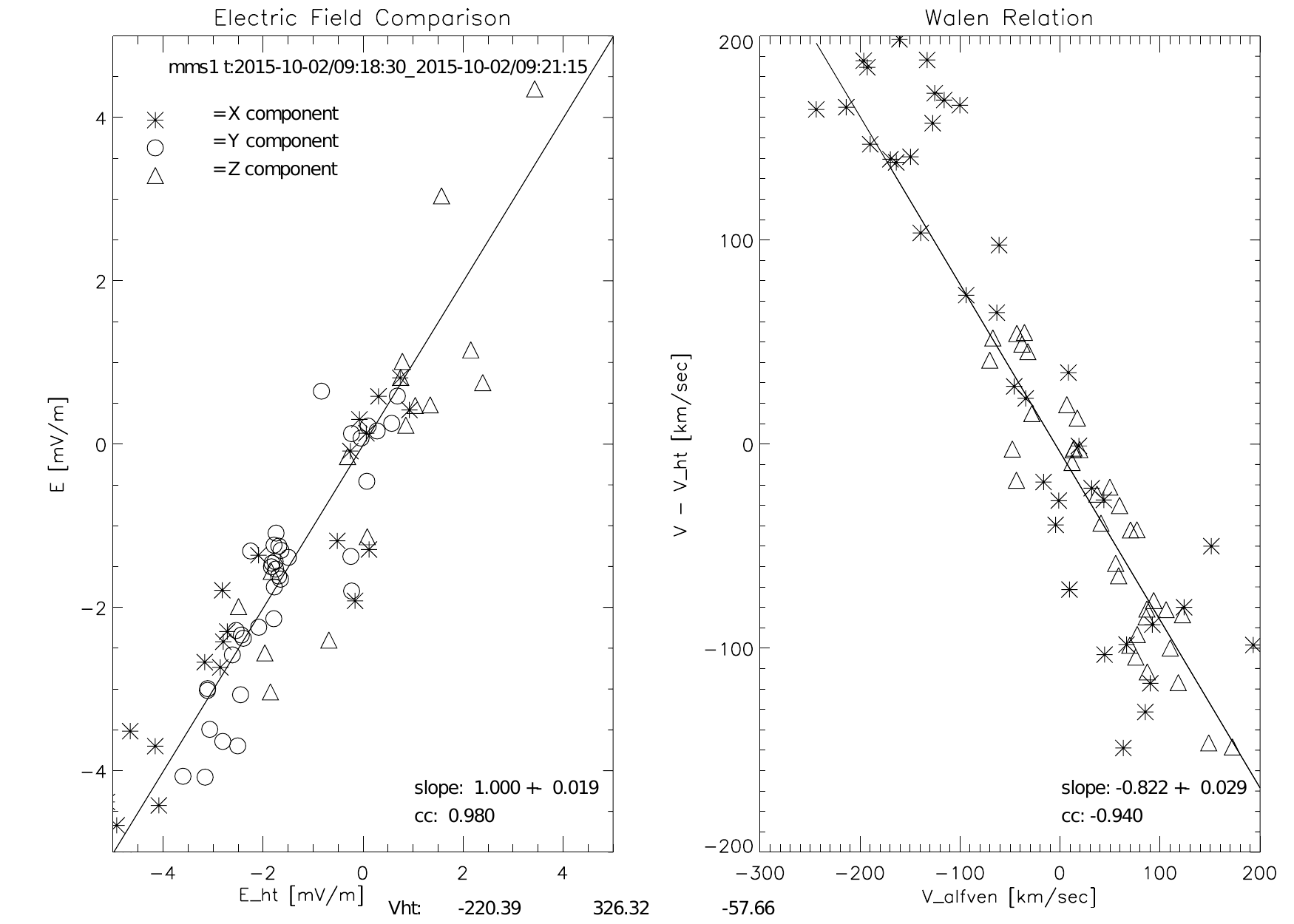}

% when using dvips, use .eps file:
% \includegraphics[width=20pc]{figsamp.eps}
%
 \caption{$\mathbf{v} \times \mathbf{B}$-Electric field comparison with the deHoffman Teller frame velocity (a), and the Wal{\'e}n relation using the pressure anisotropy correction.}
 \label{wht}
  \end{figure}

\begin{figure}[h]
\centering
% when using pdflatex, use pdf file:
\includegraphics[width=20pc]{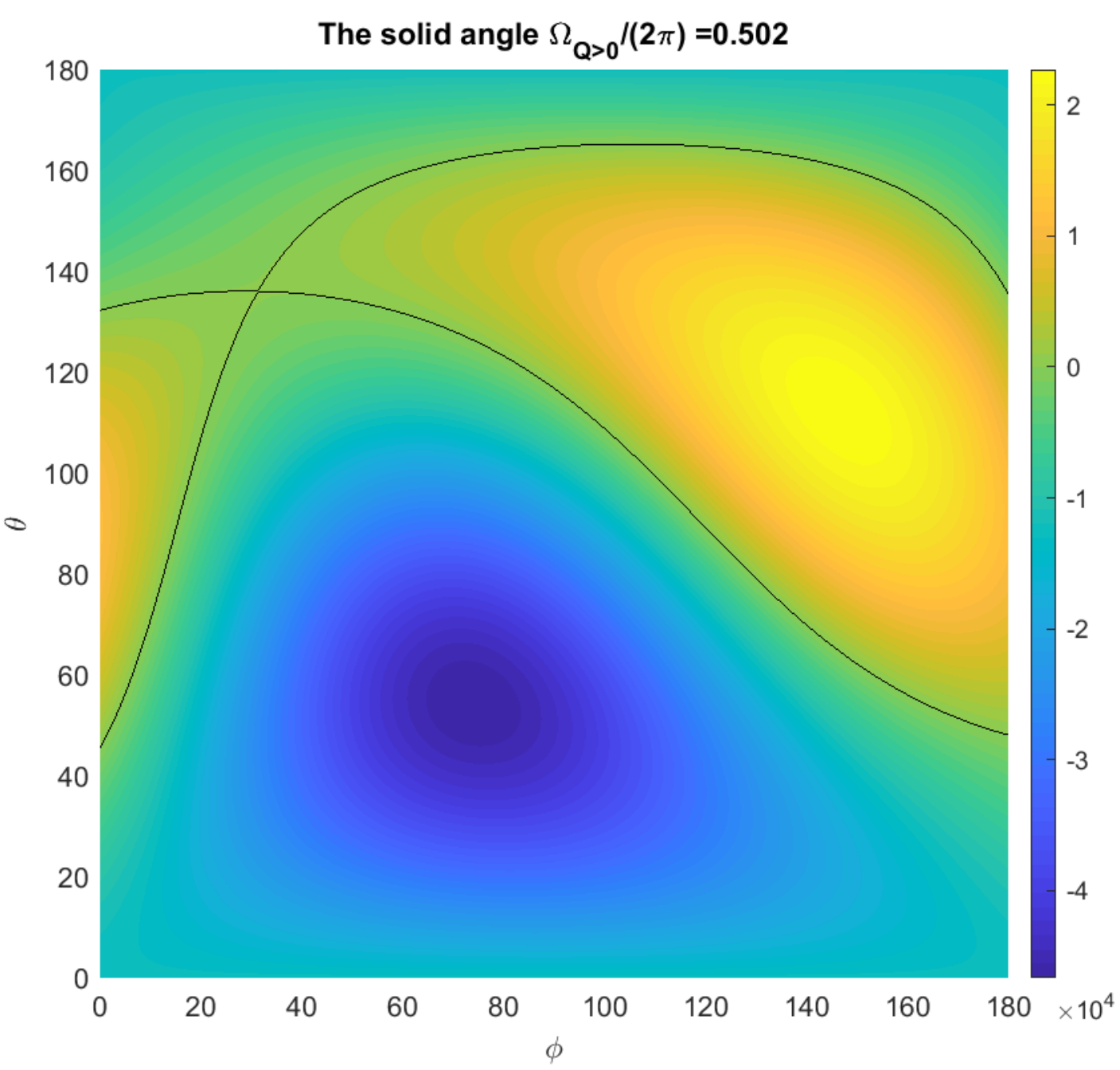}

% when using dvips, use .eps file:
% \includegraphics[width=20pc]{figsamp.eps}
%
 \caption{The value $Q$ as the function of the unit vector of the KH $k$-vector in $\theta$, $\phi $ -space of the spherical coordinates. The color code values above zero mark the $k$-vector orientations with respect to the velocity shear that satisfy the onset condition between 9:21-9:24 UT. Computation of the solid angle of the unstable orientations indicates that about 50 percent of of all possible directions are KH unstable.}
 \label{KHI}
  \end{figure}

\acknowledgments
 The work of KN is supported by NASA grant NASA grant \#NNX17AI50G. Work by XC and XM is supported by NASA grant \#NNX16AF89G.
  Authors would like to acknowledge the MMS team, especially the FGM, FPI, FEEPS and EIS instrument teams. All MMS data were downloaded through the MMS Science Data Center accessible at https://lasp.colorado.edu/mms/sdc/public/, and we recognize the efforts from all who contribute to this service. We acknowledge use of NASA/GSFC's Space Physics Data Facility's OMNIWeb (http://omniweb.gsfc.nasa.gov) service, and OMNI data. Simulation results have been provided by the Community Coordinated Modeling Center at Goddard Space Flight Center through their public Runs on Request system (http://ccmc.gsfc.nasa.gov). The SWMF/BATSRUS with RCM Model was developed by T. Gombosi et al., R. Wolf et al., S. Sazykin et al., and G. Toth et al., at the The Center for Space Environment Modeling (CSEM) at the University of Michigan.  
\listofchanges
%%%

\end{document}